\documentclass[twocolumn, prl, showpacs,superscriptaddress,floatfix]{revtex4}
\usepackage{graphicx}
\usepackage{graphicx}
\usepackage{epstopdf}

\begin{document}
\title{Quantum Criticality of 1D Attractive Fermi Gas}

\author{ Xiwen Guan}
\affiliation{Department of
Theoretical Physics, Research School of Physics and Engineering,
Australian National University, Canberra ACT 0200, Australia}
\author{ Tin-Lun Ho}
\affiliation{Department of Physics, The Ohio State
University, Columbus, OH 43210}

\date{\today}

\begin{abstract}

We obtain an analytical equation of state  for one-dimensional strongly attractive Fermi gas  for all parameter regime in current experiments. From the equation of state we derive  universal scaling functions that  control  whole  thermodynamical properties in quantum critical regimes and   illustrate physical origin of quantum criticality.  It turns out that the critical properties of the system are described by these of free fermions and those of mixtures of fermions with mass $m$ and $2m$.  We also show how these critical properties of bulk systems can be revealed from the density profile of trapped Fermi gas at finite temperatures and can be used to determine the $T=0$ phase boundaries without any arbitrariness. 

 \end{abstract}

\maketitle

\section{I. Introduction}

Quantum critical phenomena are associated with phase transitions at zero temperature as the parameters of the system are varied. They are among the most challenging problems in condensed matter, since
 quantum fluctuations couple strongly with thermal fluctuations in this regime. 
From this viewpoint, 1D integrable models exhibiting  phase
 transitions at $T=0$ are particularly valuable, as they can be solved
 exactly using Bethe Ansatz (BA) \cite{Yang-Gaudin}. 
 These exact solutions will illustrate the microscopic origin of their quantum criticality and provide unique  advance in the study of  quantum critical phenomena and universal scaling theory at quantum criticality.  On the other hand, recent advances in cold atoms experiments have provided  a highly controlled environment to study 1D quantum gases in practically all physical regimes, thus allowing one to study the predictions of Bethe Ansatz.

Despite much work on 1D Fermi gases  \cite{Takahashi,1D-Hubbard}, there are little studies of quantum
 criticality using BA  solutions.  This is mainly because the thermodynamic properties of integrable models at finite temperature are notoriously difficult and present  formidable challenges in theoretical and mathematical physics.  On the other hand, the bosonization field theory prediction \cite{Giamarchi} on the low temperature thermodynamics of 1D many-body systems merely relies on the low-lying excitations which do  not give  proper thermal potentials in quantum critical regime.   Here, we first present these
 studies of quantum criticality and scaling functions  for 1D Fermi gas with attractive $\delta$-function
 interaction, which is known to have many interesting phases at $T=0$ \cite{Orso,Hu,Guan,Wadati,Mueller,Feiguin,Kakashvili},
 including the so-called Fulde-Ferrell-Larkin-Ovchinnikov (FFLO) \cite{FFLO}  like phase.
 Our purpose is multifold: {\bf (I)} We present exact analytic expressions of the pressure, equation of state, and other thermodynamic quantities
 in the strongly interacting regime which are valid over all the parameter regime in current experiments \cite{Hulet}.
 These expressions will be useful to compare with current results and to examine  the features of the model confirmed in the experiment \cite{Hulet}.
 {\bf (II)} We show that the pressure $P$ takes on an intuitive form. It can be regarded as a sum of the pressures of free fermions and hard core bosons, plus interactions of clusters of these entities -- 2-clusters, 3-clusters, etc.
 {\bf (III)}  We show that at criticality, the pressure of this system reduces to that of free fermions or mixtures of them. The criticality of the FFLO phase when entered from different phase is characterized by the latter.   {\bf (IV)} We show that using the scaling properties of density of each spin component in the quantum critical region,  one can map out the
 $T=0$ phase diagram of a bulk systems using the $T>0$ density data of a trapped gas, as pointed out recently by one of us
 \cite{ZhouHo}. We show how quantum criticality can help determine the presence of the FFLO like phase, universal Tomonaga-Luttinger liquid (TLL),  and how the entire $T=0$ phase diagram of the bulk system can be mapped out from the finite temperature non-uniform density profile of different spin components in experiments.  This new approach to quantum criticality   will open to  study of quantum critical phenomena of spinor Fermi and Bose  gases with higher spin  symmetries.

\begin{figure}[t]
{{\includegraphics [width=1.0\linewidth,angle=-0]{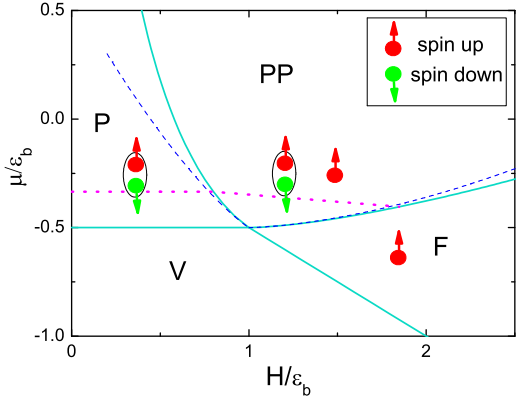}}}
\caption{(Color online)Phase diagram in $\mu-H$ plane: The phase boundaries are represented by solid lines.
The $V-F$, $V-P$ boundaries are given in Eq.(\ref{CR-1-2}), and $F-PP$ and $P-PP$ boundaries are numerical plot of Eq. (\ref{F-PP}) and (\ref{P-PP}). The blue-dashed lines are the extrapolation of the phase boundaries  Eq. (\ref{CR-S-3}) and (\ref{CR-S-4})
  in the strongly coupling regime, which is the region below the purple dotted lines, given by
  $\sqrt{\tilde{\mu}+1/2} \sim 1$, and   $\sqrt{\tilde{\mu}+h/2 } \sim 1$.
  Quantum phase transition occurs as the driving parameters $\tilde{\mu}$ and $h$ cross
  the phase boundaries at zero temperature.  The $\mu-h$  phase diagram was found from the ground state Fredholm equations in  \cite{Orso,Hu}. }
  \label{fig:Phase_diagram}
\end{figure}

\section{ II. Exact $T=0$ phase  boundaries of attractive 1D Fermi gas }


The Hamiltonian of a 1D Fermi gas with attractive $\delta$-function interaction is ${\cal K}={\cal  H}-\mu N - h M$,
\begin{equation}
{\cal H}= \frac{\hbar^2}{2m}\int {\rm d}x\left(
\partial\psi^{\dagger}_{\sigma}\partial \psi^{}_{\sigma} + c
\psi^{\dagger}_{\uparrow} \psi^{\dagger}_{\downarrow}
\psi^{}_{\downarrow} \psi^{}_{\uparrow}  \right),
\label{H}
\end{equation}
$N=\sum_{\sigma} \int {\rm d} x \psi^{\dagger}_{\sigma} \psi^{}_{\sigma}$, $M=\frac{1}{2}\int {\rm d} x (\psi^{\dagger}_{\uparrow} \psi^{}_{\uparrow} -  \psi^{\dagger}_{\downarrow} \psi^{}_{\downarrow} )$, where $\sigma=\uparrow, \downarrow$ are the spin labels.  $c$ has the dimension of (length)$^{-1}$.  Because of attractive interaction, i.e. $c< 0$,  fermions with opposite spin can form a bound pair with binding energy $\varepsilon_{b}= \hbar^2 c^2/4m$ and spatial extent $|c|^{-1}$.  The parameter $\gamma = |c|/n$, which is the ratio between inter-particle spacing ($1/n)$ and width of the bound pair $|c|^{-1}$, divides the system into a regime of tightly bound pairs ($\gamma<-1$), and overlapping pairs ($-1<\gamma< 0$).  For systems with spin polarization, both bound pairs and unpaired fermions coexist. The equilibration between them, even for very
small polarization, will make $h\sim \varepsilon_{b}$; and we shall later consider temperature regimes, $T< \varepsilon_{b}, h$. For our discussions, it is convenient to use dimensionless quantities where energy and length are measured is in units of $\varepsilon_{b}$ and $c^{-1}$ respectively. We shall define  $\tilde{\mu}\equiv \mu/\varepsilon_{b}$, $h\equiv H/\varepsilon_{b}$, $t\equiv T/\varepsilon_{b}$, $\tilde{n}\equiv n/|c|=\gamma^{-1}$, $\tilde{p} \equiv P/|c\varepsilon_{b}|$.  From the BA equation, it can be shown that the strongly coupling limit $\gamma>>1$ corresponds to $\sqrt{\tilde{\mu} + h/2} <<1 $ and $\sqrt{\tilde{\mu} +{1}/2} <<1$.

1D interacting Fermi gas with contact interaction is one of the most important exactly solvable  quantum many-body systems, and the associating BA solution is among the greatest  achievements in mathematical physics \cite{Yang-a}. This model contains only one interaction parameter $c$. Yet it is enough to generate a great diversity of intricate collective phenomena, including spin-charge separation (for $c>0$) and oscillatory paring $(c<0)$. The BA  solution, however, is formidable.  Despite the considerable analytical effort in deriving the thermodynamic Bethe Ansatz (TBA) equations that describe the thermodynamics of the system, they have only been solved numerically in most cases \cite{Mueller,Kakashvili,Hulet} and analytically for special  coupling regimes \cite{Guan,Erhai}.

The $T=0$ phase diagram has been worked out by Orso \cite{Orso} and by others \cite{Hu,Guan,Wadati}  using BA equations, which describe the ground state within a canonical ensemble \cite{Yang-Gaudin}.  Here we study the phase diagram  by taking the $T\rightarrow 0$ limit of the TBA  equations \cite{Takahashi} in a grand canonical ensemble \cite{Y-Y} and present analytical critical fields in $\mu-H$ plane.
The phase diagram is shown in Figure~ \ref{fig:Phase_diagram}.  It consists of four phases:
vacuum $(V)$, fully paired phase $(P)$, ferromagnetic phase $(F)$, and
partially paired $(PP)$ or ($FFLO$-like) phase. The phase boundaries between $V-F$, $V-P$, $F-PP$, and $P-PP$ are denoted as $\mu_{c1}$ to $\mu_{c4}$ respectively. 
The quantum phase transitions in the 1D
attractive Fermi gas are determined by the following dressed energy
equations \cite{Takahashi,Guan}
\begin{eqnarray}
\epsilon^{\rm b}(\Lambda)&=&2\left(\Lambda^2-\mu
-\frac{c^2}{4}\right)-\int_{-B}^{B}a_2(\Lambda-\Lambda'){\epsilon^{\rm
    b}}(\Lambda')d\Lambda' \nonumber\\ &
&-\int_{-Q}^{Q}a_1(\Lambda-k){\epsilon^{\rm u}}(k)d k,\nonumber\\
\epsilon^{\rm u}(k)&=&\left(k^2-\mu
-\frac{H}{2}\right)-\int_{-B}^{B}a_1(k-\Lambda){\epsilon^{\rm
    b}}(\Lambda)d\Lambda
\label{TBA-F}
\end{eqnarray}
which are obtained from the TBA equations  in the
limit $T\to 0$.  In  the above equations $a_m(\lambda)=\frac{1}{2\pi}\frac{m|c|}{(mc/2)^2+\lambda ^2}$. The dressed energy $\epsilon^{\rm b}(\Lambda)\le 0$
($\epsilon^{\rm u}(k)\le 0$) for $|\Lambda|\le B$ ($|k|\le Q$) correspond
to the occupied states. The positive part of $\epsilon^{\rm b}$
($\epsilon^{\rm u}$) corresponds to the unoccupied states.
The integration boundaries $B$ and $Q$ characterize the Fermi surfaces for
bound pairs and unpaired fermions, respectively.
The phase boundary can be worked out by analyzing the band fillings
with respect to the field $H$ and the chemical potential $\mu$ at
$T=0$.
The ($V-F$) phase boundary is determined by the conditions $\epsilon^{\rm u}(0)\le 0 $ and $\epsilon^{\rm b} (0)> 0$, which gives $\mu_{c1}=-h/2$.  Whereas the ($V-P$) phase boundary is determined by the conditions  $\epsilon^{u}(0)> 0 $ and $\epsilon^{b} (0)\le 0$, which gives $\mu_{c2}=-1/2$. 

The ($F-PP$) phase boundary  is  determined by $\epsilon^{\rm b}(0)\le 0$ and $\epsilon^{\rm u}(\pm Q)=0$ which gives   the  critical field in dimensionless units by  
\begin{equation}
\mu_{c3}=-\frac{1}{2}-\frac{1}{2\pi}\left[ Q-(2\mu_{c3}+h+1)\arctan{Q}\right]
\label{F-PP}
\end{equation}
with $Q=\sqrt{2\mu_{c3}+h}$ \cite{footnote}, see the solid ($F-PP$) phase  line $\mu_3$ in Fig.~\ref{fig:Phase_diagram}.
The most complicated phase boundary indicating the quantum phase transition ($P-PP$) from
fully paired phase into the partially polarized phase may be
determined by the conditions $\epsilon^{\rm u}(0)\le 0$  and
$\epsilon^{\rm b}(\pm B) =0$, i.e. the Fermi sea of unpaired fermion
starts filling.  Thus we have
\begin{eqnarray}
\mu_{c4}&=&-\frac{h}{2}-\frac{4}{\pi}\int_{-\tilde{B}}^{\tilde{B}}\frac{ \epsilon^b(\Lambda)d\Lambda }{1+4\Lambda^2}  \nonumber
\\
\epsilon^b(\Lambda)&=&2\Lambda^2-\mu_{c4}-\frac{1}{2}-\frac{1}{\pi}\int_{-\tilde{B}}^{\tilde{B}}\frac{ \epsilon^b(\Lambda')d\Lambda'
 }{1+(\Lambda-\Lambda')^2} \nonumber
\\ \tilde{B}^2&=&\frac{1}{2}\left(\mu_{c4}+\frac{1}{2} \right)
+\frac{1}{2\pi}\int_{-\tilde{B}}^{\tilde{B}} \frac{\epsilon^b(\Lambda)d\Lambda }{1+(\tilde{B}-\Lambda)^2},\label{P-PP}
\end{eqnarray}
which provide exact critical field $\mu_{c4}$, see  solid $P-PP$  line in Fig.~\ref{fig:Phase_diagram}.  In order to study quantum criticality of strongly Fermi gas, close forms of critical fields are essential to determine scaling functions of thermodynamical properties. They are 
\begin{eqnarray}
\mu_{c1}&=&-\frac{h}{2}; 
\,\,\,\,
\mu_{c2}=-\frac{1}{2}, \label{CR-1-2}\\
\mu_{c3}&=&
 -\frac{1}{2}\left(1-\frac{2}{3\pi}(h-1)^{\frac{3}{2}}-\frac{2}{3\pi^2}(h-1)^2\right),\label{CR-S-3}\\
 \mu_{c4}&=&
-\frac{h}{2}+\frac{4}{3\pi}(1-h)^{\frac{3}{2}}+\frac{3}{2\pi^2}(1-h)^2. \label{CR-S-4}
\end{eqnarray}
While Eq.(\ref{CR-1-2}) applies to all regimes, Eq.(\ref{CR-S-3}) and (\ref{CR-S-4}) are expressions in the strongly interacting regime, see the dashed lines in Fig.~\ref{fig:Phase_diagram}.  The above critical fields (\ref{CR-S-3}) and (\ref{CR-S-4}) can  be also obtained  by converting the critical fields in the $h-n$  plane, which were found in \cite{Guan}, into the $\mu-h$  plane. 

\section{III. Equation of state}

The lack of analytic solutions of the TBA equation has made calculations of physical properties cumbersome, severely limits ones ability to make predictions and to identify the physical origin of observed effects.  While bosonization or Luttinger liquid theory can provide qualitative information of low temperature properties, they do not give equation of states and are accurate only within  a limited range of temperature. The construction of more transparent solutions for thermodynamic functions becomes even more desirable in view of the recent progress in the experiments on 1D Fermi gas with attractive interaction, as well as  the successes in deducing equation of state of Fermi gases from non-uniform density data. A close analytic form of thermodynamic functions will certainly make comparisons between theory and experiments easier.

The thermodynamics of this system has been studied analytically \cite{Guan,Erhai} and numerically \cite{Mueller,Kakashvili,Hulet} using TBA equations. Ref. \cite{Guan} studies the free energy at very low temperatures for the spin balanced case, and shows that it is well described by Tomonaga-Luttinger liquid theory. The Luttinger description, however, is incapable of describing quantum criticality for it does not contain the critical  fluctuations.
TBA equations are a  complex set of equations in terms of the so-called ``dressed energies" of bound pairs, unpaired fermions, and spin wave bound states. In Ref.\cite{Erhai}, these equations were recasted into coupled equations of thermodynamic quantities, obtained by approximating the dressed energies to the lowest order in $t$, and taking the $T|c|\rightarrow \infty$ limit for the so-called ``string contributions" which describe the effect of the spin waves.    To describe quantum criticality accurately, however, higher order terms in $t$ in the dressed energy have to be retained. (See footnote \cite{note} below).
From the TBA equations \cite{Takahashi,Guan}, it can be shown that the dressed energies can be calculated in terms of plolylog functions 
\begin{eqnarray}
&&\epsilon^{\rm b}(k)= 2(\frac{h^2}{2m}k^2-\mu-\frac{\hbar^2}{2m}\frac{c^2}{4})+\frac{|c|p^b}{c^2+k^2}\nonumber\\
&& +\frac{T^{\frac{5}{2}}}{4\sqrt{2\pi}|c|^3}\mathrm{Li}_{\frac{5}{2}}\left(-e^{\frac{A_0^{\rm
      b}}{T}}\right)+\frac{|c|p^u}{\frac{c^2}{4}+k^2}\nonumber \\
      &&+\frac{T^{\frac{5}{2}}}{\sqrt{\pi}|c|^3}\mathrm{Li}_{\frac{5}{2}}\left(-e^{\frac{A_0^{\rm
      u}}{T}}\right)+O\left(\frac{1}{|c|^4}\right)\\
  && \epsilon^{\rm u}(k)=    \frac{\hbar^2}{2m}k^2-\mu-\frac{H}{2}+\frac{p^{\rm b}}{2}\frac{|c|}{\frac{c^2}{4}+k^2}\nonumber\\
  &&
  +\frac{\sqrt{2}}{\sqrt{\pi}}\frac{1}{|c|^3}\frac{T^{\frac{5}{2}}}{\left(\frac{\hbar^2}{2m}\right)^{\frac{3}{2}}}\mathrm{Li}_{\frac{5}{2}}\left(-e^{\frac{A_0^b}{T}}\right)\nonumber\\
  &&-Te^{-\frac{H}{T}}e^{-K}I_0(K)+O\left(\frac{1}{c^4},e^{\frac{-2H}{T}}\right)
\end{eqnarray}    
where 
\begin{eqnarray}
A_0^{\rm b}&\approx &2\mu +\frac{c^2}{2}-\frac{p^{\rm b}}{|c|}-\frac{4p^{\rm
    u}}{|c|},\nonumber\\
    A_0^{\rm u}&\approx &\mu +\frac{H}{2}-\frac{2p^{\rm
    b}}{|c|}-\frac{4p^{\rm u}}{|c|} \nonumber
\end{eqnarray}    
and $Li_{s}(z) = \sum_{k=1}^{\infty}z^{k}/k^{s}$ is the polylog function,
$K=-\frac{t^{\frac{1}{2}}}{\sqrt{2\pi}}f_{3/2}^{u}$,  and $I_0(x)=\sum_{k=0}^{\infty}\frac{1}{(k!)^2}(\frac{x}{2})^{2k}$ comes from the so-called ``string" or spin wave contributions.  The above result depends on an important observation that the convolution terms in the TBA equations converge quickly as $\epsilon^{\rm b}(k)$ and $\epsilon^{\rm u}(k)$  become greater than zero at low temperatures. 
Using the above asymptotic of   dressed energies we can calculate pressure in a straightforward  way.
Substituting above dressed energies into the pressure  per unit length $p=p^b+p^u$ 
\begin{eqnarray}
p^b&=& \frac{T}{\pi}\int_{-\infty}^{\infty}dk\ln(1+\mathrm{e}^{-\epsilon^{\rm
      b}(k)/{T}})\nonumber\\
 P^u     &=&\frac{T}{2\pi}\int_{-\infty}^{\infty}dk \ln(1+\mathrm{e}^{-\epsilon^{\rm u}(k)/{T}}), \label{pressureL}
\end{eqnarray}
and taking integration by part, thus  we obtain the following dimensionless form of the pressure of the system as 
 \begin{equation}
 \tilde{p}(t, \tilde{\mu}, h)\equiv p/(|c|\varepsilon_b)=\tilde{p}^b+\tilde{p}^{u},
 \end{equation}
 where $\tilde{p}^b$ and $\tilde{p}^{u}$ can be interpreted as the pressure of the bound pair and unpaired fermions, and are coupled through the following set of equations,
\begin{eqnarray}
\tilde{p}^b&=&-\frac{t^{\frac{3}{2}}}{2\sqrt{\pi}}F_{3/2}^{b}\left[1+\frac{\tilde{p}^b}{8}+ 2\tilde{p}^u\right]+O(t^4) \label{Pres-p}\\
\tilde{p}^u&=&-\frac{t^{\frac{3}{2}}}{2\sqrt{2\pi}}F_{3/2}^{u}\left[1+ 2\tilde{p}^b\right]+O(t^4) \label{Pres-u}\\
\frac{X_b}{t}&=&\frac{\nu_b}{t}-\frac{\tilde{p}^b}{t}-\frac{4\tilde{p}^u}{t}-\frac{t^{\frac{3}{2}}}{\sqrt{\pi}}\left( \frac{1}{16}f_{5/2}^{b}+\sqrt{2}f_{5/2}^{u}\right)\label{Xb}\\
\frac{X_u}{t}&=&\frac{\nu_u}{t}-\frac{2\tilde{p}^b}{t}-\frac{t^{\frac{3}{2}}}{2\sqrt{\pi}}f_{5/2}^{b}+e^{-\frac{h}{t}}e^{-K}I_0(K)
\label{dress-X}
\end{eqnarray}
where the functions $F_n^b$, $F_n^u$, $f_n^b$, and $f_n^u$ are defined as
\begin{eqnarray}
&& F_n^b \equiv Li_n\left(-e^{\frac{X_b}{t}}\right),\qquad F_n^u \equiv Li_n\left(-e^{\frac{X_u}{t}}\right) \nonumber \\
&&f_n^b\equiv Li_n\left(  -e^{\frac{\nu_b}{t}}\right),\qquad f_n^u\equiv Li_n\left(-e^{\frac{\nu_u}{t}}\right)
\end{eqnarray}
where $\nu_{b}\equiv 2{\tilde{\mu}}+1$, $\nu_{u}\equiv  \tilde{\mu} + h/2$;  $X_b$, $X_u$ are defined as in Eq.(\ref{Xb}) and Eq.(\ref{dress-X}).

\begin{figure}[t]
{{\includegraphics [width=0.90\linewidth]{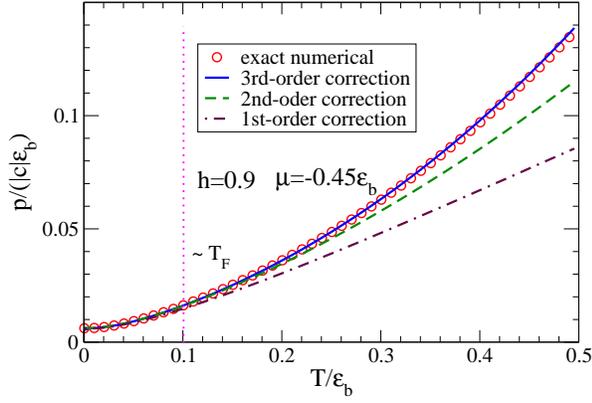}}}
\caption{(Color online)Pressure vs temperature: Comparisons between the numerical results from Eq.(\ref{Pres-p}) to  (\ref{dress-X})  (represented by circles) with the analytical result from Eq.  (\ref{SOE}) up to the order
  $Y^{(u,b)}_{\frac{1}{2}}$ (1st order), $Y^{(u,b)}_{1}$ (2nd order)
  and $Y^{(u,b)}_{\frac{3}{2}}$ (3rd oder). }\label{pressure}
\end{figure}

The derivation of Eq.(\ref{Pres-p}) to  (\ref{dress-X}) is very involved.  That we present these equations is to prepare for the later discussions of the mathematical manipulation needed to extract the singularity near the quantum critical point.  To solve these equations, one substitutes  Eqs (\ref{Xb}) and (\ref{dress-X}) into Eq.(\ref{Pres-p}) and (\ref{Pres-u}). This gives two coupled equations of $\tilde{p}^b$ and $\tilde{p}^u$, which can be solved by iteration. From Eqs (\ref{Xb}) and (\ref{dress-X}), we can rewrite the functions $\mathrm{Li}_{\frac{3}{2}}\left(-e^{X_b/t}\right)$ and $\mathrm{Li}_{\frac{3}{2}}\left(-e^{X_u/t}\right)$ in terms of the function $f_n^b$ and $f_n^u$. After lengthy algebra, the pressures are given by $\tilde{p}=\sqrt{2} \tilde{p}^{\rm b} +\tilde{p}^{\rm u}$ with 
\begin{eqnarray}
\tilde{p}^x &=& -\frac{t^{\frac{3}{2}}}{2\sqrt{2\pi}}\left[
  f_{\frac{3}{2}}^x+t ^{\frac{1}{2}}Y_{\frac{1}{2}}^x+ t Y_{1}^x+t^{\frac{3}{2}}Y^x_{\frac{3}{2}}+O(t^{2})  \right].\label{pressure-0}
 \label{SOE}
\end{eqnarray}
where $x=b,u$ and $\left\{ Y_n^b\right\}$ and  $\left\{ Y_n^u\right\}$ are given by
\begin{eqnarray}
Y_{\frac{1}{2}}^b&=&\frac{1}{\sqrt{\pi}}
f_{\frac{1}{2}}^b\left(\frac{1}{2}f_{\frac{3}{2}}^b+\sqrt{2}
f_{\frac{3}{2}}^u\right),\qquad Y_{\frac{1}{2}}^u=\frac{1}{\sqrt{\pi}} f_{\frac{1}{2}}^uf_{\frac{3}{2}}^b,\nonumber\\
Y_{1}^b&=&\frac{1}{\pi} \left[f_{\frac{1}{2}}^b\left(\frac{1}{4}f_{\frac{1}{2}}^bf_{\frac{3}{2}}^b+\sqrt{2}f_{\frac{1}{2}}^u
f_{\frac{3}{2}}^b+ \frac{1}{\sqrt{2}}f_{\frac{1}{2}}^b
f_{\frac{3}{2}}^u\right) \right. \nonumber \\
  &  &
\left. + f_{-\frac{1}{2}}^b
 \left(
 \left(f_{\frac{3}{2}}^u\right)^2+\frac{1}{8}\left(f_{\frac{3}{2}}^b\right)^2+\frac{1}{\sqrt{2}}f_{\frac{3}{2}}^b
f_{\frac{3}{2}}^u\right)\right],
\nonumber\\
Y_{1}^u&=&\frac{1}{\pi}\left[\sqrt{2}f_{\frac{1}{2}}^bf_{\frac{1}{2}}^uf_{\frac{3}{2}}^u+\frac{1}{2}f_{\frac{1}{2}}^bf_{\frac{1}{2}}^uf_{\frac{3}{2}}^b+\frac{1}{2}\left(f_{\frac{3}{2}}^b\right)^2f_{-\frac{1}{2}}^u\right].\label{Y_1_2}
\end{eqnarray}
The expressions of $Y^{b}_{\frac{3}{2}}$ and $Y^{u}_{\frac{3}{2}}$ are given in {\em Appendix}.
The accuracy of Eq.(\ref{SOE}) is shown in Figure 2, where the result of the expansion is compared with the  numerical solution of the recast TBA equations.  Eq.(\ref{Pres-p}) to  (\ref{dress-X}) served as equation of state (represented by circles). The vertical dotted line represents the Fermi temperature $T_{F}$ in the Rice experiment in ref. \cite{Hulet}, where $T_{F}/\epsilon_{b}\sim 0.1$.
Figure 2 shows that for  $t<0.1$, (corresponds to $T<T_{F}$ in \cite{Hulet}), the pressure is accurately given by the first order correction  in Eq.(\ref{SOE}) (to about 1$\%$). At higher temperatures, higher order terms are needed. It is worth noting that by including terms up to  third order,  Eq.(\ref{SOE}) agree essentially with the exact result for $t>0,5$.

Furthermore, if defining 
\begin{eqnarray}
\tilde{p}^{u}_{o} &\equiv&  -\frac{t^{\frac{3}{2}}}{2\sqrt{2\pi}}f_{\frac{3}{2}}^u, \qquad \tilde{p}^{b}_{o} \equiv  -\frac{t^{\frac{3}{2}}}{2\sqrt{\pi}}f_{\frac{3}{2}}^b,\nonumber\\
 \tilde{n}^{(b,u)}_{o} &=&\frac{ \partial \tilde{p}^{(b,u)}_{o}}{\partial \tilde{\mu}^{(b,u)}}, \qquad  \tilde{\kappa}^{(b,u)}_{o}= \frac{\partial \tilde{n}^{(u,n)}_{o}}{\partial \tilde{\mu}^{(b,u)}},
 \end{eqnarray}
 up to the order of $Y^{(u,b)}_{1}$,  we can rewrite the pressure (\ref{pressure-0}) as 
\begin{eqnarray} 
\tilde{p}^{b}&\approx & \tilde{p}_{o}^{b} - \tilde{n}_{o}^{b}\left(\tilde{p}^{b}_{o}+4\tilde{p}_o^u\right) +\tilde{n}_o^b\left(\tilde{n}_o^b\tilde{p}^b_o+8\tilde{n}_o^u\tilde{p}_o^b+4\tilde{n}_o^b\tilde{p}^u_o\right)\nonumber\\
&&+\tilde{k}_o^b\left(8(\tilde{p}^u_o)^2+\frac{1}{2}(\tilde{p}^b_o)^2 +4\tilde{p}^b_o\tilde{p}^u_o\right),\nonumber\\ 
 \tilde{p}^{u}&\approx & \tilde{p}_{o}^{b} -2\tilde{n}_o^u\tilde{p}_o^b+8\tilde{n}_0^b\tilde{n}_o^u\tilde{p}^u_o+2\tilde{n}_o^b\tilde{n}_o^u\tilde{p}_o^b+2\tilde{k}_o^b(\tilde{p}^u_o)^2
 \end{eqnarray}
which give insight into understanding cluster-cluster  effect. From these analytical result of pressures, we see that the bound pairs  interfere with each other and with excess fermions.  The $Y$-terms in Eq.(\ref{SOE}) can be viewed as interactions of boson-boson and boson-fermion clusters of  increasingly size. We find  that
The pressure Eq. (\ref{SOE}) can reach $95\%$ accuracy for  $T/\varepsilon_{b}=0.2$ by including upto the $Y^{x}_{1}$ terms, and has
less than $1\%$ error if one includes the $Y_{\frac{3}{2}}^{x}$ terms, see  Figure~\ref{pressure}.

Finally, we note that in the limit $|c|\rightarrow\infty$, the pressure (\ref{pressureL})  gives the following equation of state 
$P(T,\mu,H)=\sqrt{2}{\cal P}(2\mu_b)+ {\cal P}(\mu_u)$ with $T\ll  \mu_b, \,\mu_u$,
 where 
 \begin{equation}
 {\cal P}(x)\approx \frac{2\sqrt{2}}{3\pi}\sqrt{\frac{m}{\hbar^2}}x^{\frac{3}{2}}\left(1+\frac{\pi^2}{8}\left(\frac{T}{x}\right)^2+\frac{7\pi^4}{640}\left(\frac{T}{x}\right)^4 \right)
 \end{equation}
  is  the pressure of a 1D Fermi gas with mass $m$.  In this limit,   for fixed total density $n$, the effective
chemical potentials for pairs and unpaired fermions are given by
\begin{equation}
\mu_{\rm b} \approx  \frac{\hbar^2n^2\pi^2}{2m}\frac{(1-P)^2}{16},\label{Mu-b}
\end{equation}
\begin{equation}
\mu_{\rm u} \approx \frac{\hbar^2n^2\pi^2}{2m}P^2.\label{Mu-u}
\end{equation}
Where $P=(N_{\uparrow}-N_{\downarrow})/(N_{\uparrow}-N_{\downarrow})$ is the polarization.
This result reflects the fact that the system  reduces to a mixture of hard core boson (mass $2m$) and a gas of unpaired fermions (mass $m$) in this limit. Since hard core bosons in 1D behaves like fermions, the thermodynamics of the system is that of a mixture of two Fermi gases with mass $2m$ and $m$. 

\section{IV. Quantum Criticality}

\begin{figure}[t] 
{{\includegraphics [width=0.90\linewidth]{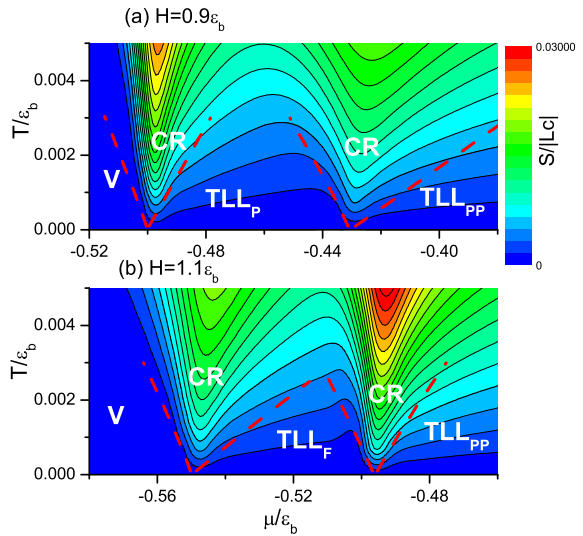}}}
\caption{(Color online)Contour plot  entropy $S$  vs  chemical potential $\mu$  from Eq. Eq.(\ref{Pres-p}) to  (\ref{dress-X})  for  different values of effective magnetic field \cite{note-figure}.  The cross-over temperatures separated TLL phases from quantum critical regimes are determined from the breakdown of linear temperature-dependent entropy.  Here $TLL_F$ stands for the Tomonaga-Luttinger liquid of the single component  of unpaired fermions. While $TLL_{P}$ stands for the  TLL of pairs and   $TLL_{PP}$ for the two-component TLL of  FLLO-like states.  The left-most dashed line separates quasi-classical   regime from the  quantum critical regime. The cross-over temperatures merge at  critical points as $T\to 0$.  }\label{fig:s}  
\end{figure}

 Quantum critical behaviors are reflected in the singularities in thermodynamic quantities such as
 density $\tilde{n}=n/|c|$, compressibility $\tilde{\kappa}=\partial \tilde{n}/\partial \tilde{\mu}$, and magnetization $\tilde{M}=M/|c|$, which are derivatives of pressure $p$ with respect to $\mu$ and $h$.
 Such singularities, however, cannot be obtained by directly differentiating the expansion Eq.(\ref{SOE}). The reason is that even though Eq.(\ref{SOE}) gives highly accurate value for the pressure with a few terms, all higher order terms which contribute  insignificantly to the pressure have singularities when differentiated with respect  to $\mu$.  To account for these singular structures, say, in the density, one must first differentiate Eq.(\ref{Pres-p})-(\ref{dress-X})  with respect to $\tilde{\mu}$ to obtained coupled equations of  $\partial \tilde{p}^{x}/\partial \tilde{\mu}$  and then solves for these quantities by iteration.  Quantum phase transition occurs as the driving parameters $\tilde{\mu}$ and $\tilde{h}$ cross the phase bounaries at zero temperature.   At very low temperatures, $T\ll \varepsilon_b$
(Boltzmann's constant $k_B=1$), the thermodynamics of the 1D FFLO like 
phase is governed by the linearly dispersing phonon modes, i.e., the
long wavelength density fluctuations of the two weakly coupled gases.
In this low temperature regime, spin strings are suppressed.  The
suppression of spin fluctuations leads to a universality class of a
two-component TLL  in the gapless phase of FFLO, where the
charge bound states form hard-core composite bosons.  The leading low
temperature corrections to the free energy give \cite{Guan,Erhai}
\begin{equation}
f \approx f_0-\frac{\pi
  T^2}{6\hbar}\left(\frac{1}{v_{b}}+\frac{1}{v_u}\right). \label{FreeE}
\end{equation}
Here $f_0$ is the ground state energy and $v_{b,u}$ are the sound velocities of pairs and unpaired
fermions, i.e. 
\begin{eqnarray}
v_{\rm b} &\approx &\frac{v_{\rm F}(1-P)}{4}
\left(1+\frac{(1-P)}{|\gamma|} +\frac{4P}{|\gamma|}\right),\nonumber\\
v_{\rm u}&\approx & v_{\rm F} P\left( 1 +
\frac{4(1-P)}{|\gamma|}\right).
\end{eqnarray}
Here the Fermi velocity is $v_{\rm F}=\hbar \pi n/m$. 
The TLL  maintains below the cross-over
temperature at which  the relation of the linear
temperature-dependent  entropy (or specific) heat breaks down. The equation of state Eq.(\ref{Pres-p}) to  (\ref{dress-X}) and the TLL nature (\ref{FreeE}) allow ones to quantitatively investigate quantum criticality of the Fermi gas,  see Fig.~ \ref{fig:s}.

Using the standard thermodynamic relations, we can derive close forms   of density, magnetization, compressibility
which allow ones to capture universal  low temperature thermodynamics of the Fermi gas as well as  critical phenomena.  Without losing generality, we can safely ignore spin wave contribution at quantum criticality due to its exponentially small contribution as $T\to 0$. 
For our convenience in calculating the thermodynamical properties, we denote
\begin{eqnarray}
f_n^{A_b}&\equiv &Li_n\left(  -e^{\frac{A_b}{t}}\right), \,\frac{A_b}{t}=\frac{\nu_b}{t}+\frac{t^{\frac{1}{2}}}{\sqrt{\pi}}\left( \frac{1}{2}
  f_{\frac{3}{2}}^{b}
+\sqrt{2}f_{\frac{3}{2}}^{u}\right),\nonumber\\
f_n^{A_u}&\equiv & Li_n\left(-e^{\frac{A_u}{t}}\right),\,\,\, \frac{A_u}{t}=\frac{\nu_u}{t}+\frac{t^{\frac{1}{2}}}{\sqrt{\pi}}
  f_{\frac{3}{2}}^{b}\label{A_bu}.
\end{eqnarray}
From the equations Eq.(\ref{Pres-p})-(\ref{dress-X}),  We  derived the total density $\tilde{n}=(\tilde{n}_{\uparrow}+\tilde{n}_{\downarrow})$
\begin{eqnarray}
&&\tilde{n}=-\frac{\sqrt{t}}{\sqrt{\pi }\Delta}\left\{\frac{1}{2\sqrt{2
}}f_{\frac{1}{2}}^{A_u}\left[1+\frac{3\sqrt{t}}{2\sqrt{\pi
}} f_{\frac{1}{2}}^{A_b}-\frac{47t^{\frac{3}{2}}}{16\sqrt{\pi }}
f_{\frac{3}{2}}^{A_b} \right]\right.\nonumber  \\
&&\left.+f_{\frac{1}{2}}^{A_b}\left[1+\frac{\sqrt{t}}{\sqrt{2\pi
}} f_{\frac{1}{2}}^{A_u}
-\frac{t^{\frac{3}{2}}}{\sqrt{\pi }}\left(\frac{1}{8}
f_{\frac{3}{2}}^{A_b} +\frac{3\sqrt{2}}{2}f_{\frac{3}{2}}^{A_u} \right)\right]\right\} \label{density-EOS}
\end{eqnarray}
with 
\begin{eqnarray}
\Delta&=&1-\frac{\sqrt{t}}{2\sqrt{\pi
}}f_{\frac{1}{2}}^{A_b}-\frac{t\sqrt{2}}{\pi}f_{\frac{1}{2}}^{A_b}f_{\frac{1}{2}}^{A_u}+\frac{t^{\frac{3}{2}}}{16\sqrt{\pi}}f_{\frac{3}{2}}^{A_b}.
\end{eqnarray}
The density in dimensionless scale naturally  services as the dimensionless equation of state which contain two free-fermion-like densities with singular behaviour near different  critical points. The interaction binding energy rescales temperature.   This close form of equation of state is very convenient to make fitting of experimental finite temperature density profiles of the 1D trapped gas  within local density approximation, see \cite{Hulet,note-figure}.

Similarly,  magnetization  $\tilde{M}=(\tilde{n}_{\uparrow}-\tilde{n}_{\downarrow})/2$
\begin{eqnarray}
\tilde{M}&=&-\frac{\sqrt{t}}{2\sqrt{\pi }\Delta}\left\{\frac{1}{2\sqrt{2
}}f_{1/2}^{A_u}\left[1-\frac{\sqrt{t}}{2\sqrt{\pi
}} f_{1/2}^{A_b}-\frac{31t^{\frac{3}{2}}}{16\sqrt{\pi }}
f_{3/2}^{A_b}\right] \right.\nonumber\\
&& \left.+\frac{\sqrt{t}}{\sqrt{2\pi}}f_{1/2}^{A_b}\left[ f_{1/2}^{A_u}-t
f_{\frac{3}{2}}^{A_u}\right]\right\}; \nonumber
\end{eqnarray}
 susceptibility $\tilde{\chi}=\chi \varepsilon_b /|c|$ with $ \chi =\partial M/\partial H$
\begin{eqnarray}
\tilde{\chi} &=&-\frac{1}{8 \sqrt{2\pi }\Delta^3}\left\{\frac{1}{\sqrt{t}}f_{-1/2}^{A_u}\left[1+\frac{3\sqrt{t}}{2\sqrt{\pi
}}f_{1/2}^{A_b}
+\frac{2\sqrt{2}t}{\pi}f_{1/2}^{A_b}f_{1/2}^{A_u}\right]\right.\nonumber  \\
&&\left.+\frac{2\sqrt{2}\sqrt{t}}{\pi}f_{-1/2}^{A_b} \left(f_{1/2}^{A_u}\right)^2\right\};
\end{eqnarray}
and  compressibility $\tilde{\kappa} =\kappa \varepsilon_b/|c|$ with $\kappa =\partial n/\partial \mu$
\begin{eqnarray}
\tilde{\kappa}&=&-\frac{1}{\sqrt{\pi }\Delta^3}
\left\{\frac{1}{2\sqrt{2}\sqrt{t}}f_{-1/2}^{A_u}\left[1-\frac{5\sqrt{t}}{2\sqrt{\pi
}}f_{1/2}^{A_b}
-\frac{t}{4\pi}\left(f_{1/2}^{A_b}\right)^2\right]\right. \nonumber\\
&&\left.+\frac{2}{\sqrt{t}}f_{-1/2}^{A_b}\left[1-\frac{\sqrt{t}}{2\sqrt{2\pi
}}f_{1/2}^{A_u}
-\frac{t}{4\pi}\left(f_{1/2}^{A_u}\right)^2\right]\right\} \label{compressibility}
\end{eqnarray}
can be derived in a systematic way.

\begin{figure}[t] 
{{\includegraphics [width=1.0\linewidth]{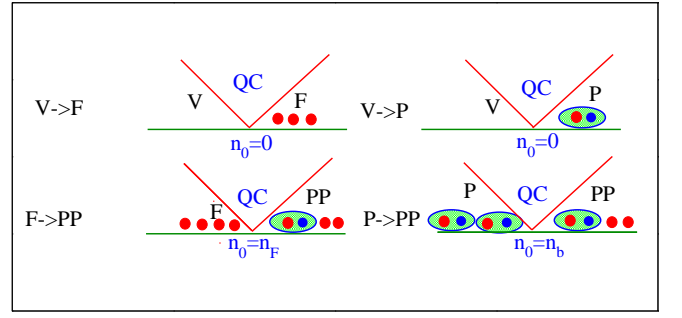}}}
\caption{(Color online)Schematic illustration  of universal critical phenomena near critical points for the 1D strongly attractive Femi gas, see Eq.  (\ref{universal}).  The density of bound pairs (excess fermions)  become a regular part as the background meanwhile  the one of excess fermions (bound pairs) become a singular part.  However, there is no back ground density for the phase transition from vacuum into the fully-paired or fully-polarized   phase,  i.e. $V\to F$ or  $V-P$. }\label{fig:schematic}  
\end{figure}

Quantum criticality describes strongly coupled thermal and  quantum fluctuations of matter as
quantum phase transitions take place at zero temperature. Such quantum
phase transitions are uniquely characterized by the critical exponents
depending only on the dimensionality and the symmetry of the system.
Here we show that scaling functions at quantum criticality  can be calculated from the above close forms of  thermodynamical properties. 
Near the critical points, we expand thermodynamical properties in the limit $|\mu-\mu_c| \ll 1$.  For the quantum critical regime, i.e. $T> |\mu-\mu_c|$,  we find  that the thermodynamical properties of the Fermi liquid of bound pairs (excess fermions)  become a regular part as the background meanwhile  the ones of the  Fermi liquid of excess fermions (bound pairs) become a singular part.   
Near the $T=0$ phase boundaries, we find that $\tilde{n}$ and $\tilde{M}$ have universal  scaling forms
 \begin{eqnarray}
(V{\rm -}F) &\,\,\,\, \,\, \tilde{n} \approx - \frac{\sqrt{t}}{2\sqrt{2\pi}} Li_{\frac{1}{2}} \left(-e^{\frac{(\tilde{\mu}-\mu_{c1})}{t}}\right) \hspace{0.5in}  \label{vfn}\\
      &  \tilde{M} \approx -\frac{\sqrt{t}}{4\sqrt{2\pi}} Li_{\frac{1}{2}} \left(-e^{\frac{(\tilde{\mu}-\mu_{c1})}{t}}\right)  \hspace{0.5in}    \label{vfm}\\
 (F{\rm -}PP) & \,\,  \tilde{n}  \approx  n_{o3} -\lambda_{1}\sqrt{t} Li_{\frac{1}{2}} \left(-e^{\frac{2(\tilde{\mu}-\mu_{c3})}{t}}\right) \hspace{0.5in}  \label{fppn}  \\
  &    \tilde{M} \approx  M_o +\lambda_{3}\sqrt{t} Li_{\frac{1}{2}} \left(-e^{\frac{2(\tilde{\mu}-\mu_{c3})}{t}}\right)\  \hspace{0.5in}  \label{fppm}
      \end{eqnarray}
      \begin{eqnarray}
 (V{\rm -}P) \hspace{0.2in}     \,\,  \tilde{n} \approx  -\frac{\sqrt{t}}{\sqrt{\pi}} Li_{\frac{1}{2} }\left(-e^{\frac{2(\tilde{\mu}-\mu_{c2})}{t}}\right) \hspace{0.5in}   \label{vpn}  \\
 \tilde{M}\approx  \frac{t}{2\sqrt{2}\pi} Li_{\frac{1}{2}} \left(-e^{\frac{2(\tilde{\mu}-\mu_{c2})}{t}}\right)Li_{\frac{1}{2} } \left(-e^{\frac{(h-1)}{2t}}\right) \sim 0   \label{vpm}\\
(P{\rm -}PP)  \hspace{0.2in}    \,\, \tilde{n} \approx  n_{o4} -\lambda_{2}\sqrt{t} Li_{\frac{1}{2}} \left(-e^{\frac{(\tilde{\mu}-\mu_{c4})}{t}}\right)  \hspace{0.3in} \label{pppn} \\
   \,\, \tilde{M}  \approx  -\frac{\sqrt{t}}{4\sqrt{2\pi}}Li_{\frac{1}{2} }\left(-e^{\frac{(\tilde{\mu}-\mu_c4)}{t}}\right)\left(1-\frac{4\sqrt{b}}{\pi}\right)  \hspace{0.3in} \label{pppm}
 \end{eqnarray}
where $n_{o3}, n_{o4}, a, b, M_{o}$ are constants independent of $\tilde{\mu}$ and $t$. The expressions are given by
 \begin{eqnarray}
 n_{o3}&=& \frac{\sqrt{a}}{2\pi}, \qquad \lambda_{1} = \frac{1}{\sqrt{\pi}} \left( 1- \frac{ 2\sqrt{a}}{\pi} +  \frac{a}{\pi^2}\right),\nonumber\\
 a&=&(h-1)(1+\frac{2}{3\pi}\sqrt{h-1}),\nonumber\\
 n_{o4}&=&\frac{ 2\sqrt{b}}{\pi}\left( 1- \frac{\sqrt{b}}{\pi} + \frac{b}{\pi^2}\right),\nonumber\\
 \lambda_2&=&\frac{1}{2\sqrt{2\pi}} \left( 1- \frac{8\sqrt{b}}{\pi} + \frac{17 b}{\pi^2}\right),\nonumber\\
  b&=&(1-h)(1+\frac{2}{\pi}\sqrt{1-h}),\nonumber \\
 M_o &=& \frac{\sqrt{a}}{4\pi}, \,\, \lambda_{3}= \frac{\sqrt{a}}{2\pi^{\frac{3}{2}}}\left(1-\frac{\sqrt{a}}{\pi}-\frac{2a}{3}\right).
\end{eqnarray}
In the above equations $n_{o3}$ and $n_{o4}$ are the background densities near the critical points $\mu_3$ and $\mu_4$, respectively. 
At quantum criticality, the   above densities can be casted into a universal scaling form,  e.g.  \cite{ZhouHo,Fisher,Sachdev,Mueller2}, 
\begin{equation}
n(\mu, T) = n_{0}+ T^{\frac{d}{z}+1-\frac{1}{\nu z}} {\cal G}\left(\frac{\mu-\mu_{c}}{T^{\frac{1}{\nu z}}}\right), \label{universal}
\end{equation}
Where the dynamic exponent $z=2$ and correlation exponent $\nu=1/2$ can be read off the scaling functions within the expressions Eq.(\ref{vfn}) to (\ref{pppm}) from which physical origin of quantum criticality is conceivable.

Eq.(\ref{vfn}) to (\ref{pppm}) has a simple interpretation.  First of all, it is straightforward to work out the critical properties of a 1D free Fermi gas (single component), as well as for a mixture of two different Fermi gases.  In the former case, the quantum phase transition (as a function of chemical potential) is from vacuum to Fermi gas,  which we denote as type ${\bf (a)}$.  In the two component case, if the two Fermi gases have different particle numbers, then there is a transition from vacuum to the majority component as a function of $\tilde{\mu}$ (for fixed $h$), i.e. type ${\bf (a)}$. There is also another transition from the majority component to a mixture, which we denote as type ${\bf (b)}$. Comparing Eq.(\ref{vfn}) to (\ref{pppm}) with the critical properties of Fermi gas mixtures, one notes that the transition $V-F$ and $V-P$ is that of type ${\bf (a)}$ where the ``fermion mass"  for the $V-F$ transition is $m$, and $2m$ for the $V-P$ transition.
In the $V-P$ case, this is due to the fact that the tightly bound fermion pair acts like a hard core boson, which in turn acts as a fermion in 1D. In contrast, the transitions $F-PP$ and $P-PP$ are of type ${\bf (b)}$, where the Fermi gas mixture is made of particles with mass $m$ and $2m$. This critical phenomena is schematically illustrated in  Fig~. \ref{fig:schematic}.

\begin{figure}[t] 
{{\includegraphics [width=0.90\linewidth]{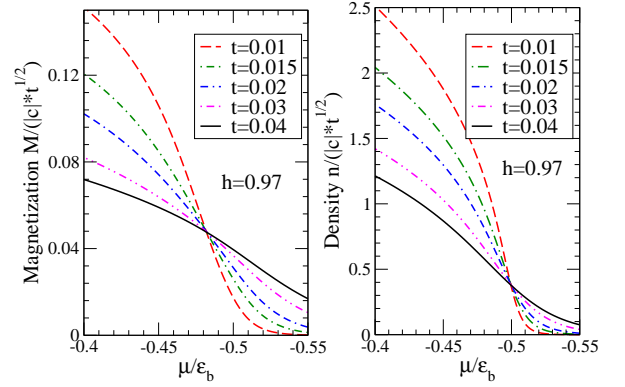}}}
\caption{(Color online)Magnetization $M$ and density $n$  vs  chemical potential $\mu$ at different temperatures.  The intersections at the right and left panel give the phase boundaries of
$V-P$ and $P-PP$ transition respectively. }\label{fig:3}  
\end{figure}

\begin{figure}[t]
{{\includegraphics [width=0.90\linewidth]{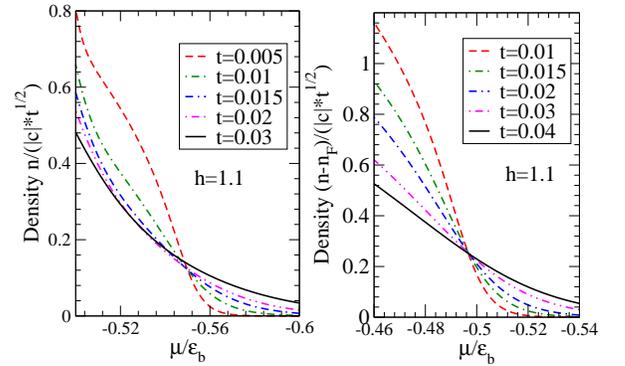}}}
\caption{(Color online)Density vs chemical potential:
 The right panel shows  the intersection of the density difference $(\tilde{n}-\tilde{n}_F)/\sqrt{t}$ at different temperatures, which gives the $F-PP$ phase boundary.  Here, we have
  $\tilde{n}_F=-\frac{\sqrt{t}}{2\sqrt{2\pi}}Li_{\frac{1}{2}}(-e^{A_u/t})$ with $A_u \approx \nu_u+t^{\frac{3}{2}}
  f_{3/2}^{b}/\sqrt{\pi}$. It  becomes the density of a free fermion system in the vicinity of $F-PP$ phase boundary, i.e. $A_u$ reduces to $\nu_u$.
 The left panel shows the  intersection of densities which gives the $V-F$ phase boundary. It also shows that the scaling form begins to fail for $t \ge 0.03$. }\label{fig:4}
\end{figure}

\section{V. Mapping out the $T=0$ phase diagram using quantum criticality}

 As pointed out in Ref.\cite{ZhouHo}, the scaling property of the density near a quantum critical point enables one to determine the {\em $T=0$} phase boundary of {\em homogenous bulk} systems from the {\em non-uniform} density profile of a {\em trapped} gas at {\em $T>0$}. All one needs to do is to plot the density profile $n(x)$ of the trapped gas as a function of local chemical potential $\mu(x) = \mu - V(x)$ at different temperatures, where $V(x)$ is the trapping potential. At low temperatures, these density curves will intersect at the same point. The chemical potential ($\mu_{c}$) associated with this intersection point  is $T=0$ phase boundary. The existence of this intersection is due to the fact that  the singular part of the density is a function of  $(\mu-\mu_{c})/T$, as seen in Eq.(\ref{vfn}) to Eq.(\ref{pppm}).
These intersections are seen in Fig.~\ref{fig:3} to Fig. ~\ref{fig:4}, which show  the numerical solution of  the recast TBA equations  Eq.(\ref{Pres-p}) to  (\ref{dress-X})  for the $n$ and $M$ as a function of  $\mu$ over a range of chemical potential much larger than the scaling region where Eq.(\ref{vfn}) to Eq.(\ref{pppm}) hold.  They are equivalent to plotting the experimental density data in a manner mentioned above.
The left and right panel of Fig.~ \ref{fig:3} show the $\tilde{\mu}$ dependence of  $\tilde{M}$ and
 $\tilde{n}$ across the $V-PP$  phase boundaries (Eq.(\ref{pppm}) and Eq.(\ref{vpn}) )
respectively.
In  Fig.~ \ref{fig:4}, the left panel shows $\tilde{n}$ vs $\tilde{\mu}$ across the $V-F$ phase boundary, Eq.(\ref{vfn}). The right panel  shows $\tilde{n}-\tilde{n}_F$ vs $\tilde{\mu}$ across  the $F-PP$ phase boundary, Eq.(\ref{fppn}). The intersections of these curves yield the critical value $\mu_{c}$ at $T=0$.  In these figures, we displayed temperature variations from $t=0.01$ to $t=0.04$.  The former corresponds to $T/T_{F}= 0.1$ in the recent Rice set up, which is the temperature in the current Rice experiment \cite{Hulet}.

In general, the slope of densities at the quantum criticality  can be written as the following universal scaling form near the critical points:
\begin{eqnarray}
\left(n(t,\mu)-n_0\right)t^{-1/2}=\lambda\sum_{m=0}^{\infty}c_m x^m 
\end{eqnarray}
where  $ x=\frac{\mu-\mu_{c}}{t}$  and the coefficients  $c_m=\frac{1}{m!}Li_{\frac{1}{2}-m}(-1)$. The intersection behaviour  can accurately determine the temperature from the slops near the intersection  point. For a given polarization (or say fixed polarization), the $\lambda$ becomes constant, (see,  Eq.(\ref{vfn}) to Eq.(\ref{pppm})).  Thus the slops are uniquely fixed by temperatures. Perhaps  this scaling behaviour is a good thermometry \cite{Mueller-note}. 

\begin{figure}[t] {{\includegraphics [width=0.90\linewidth]{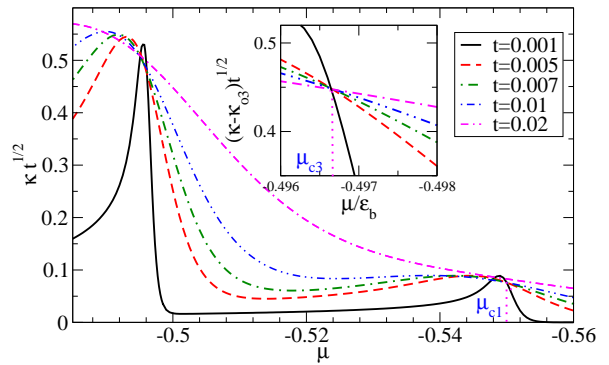}}}
 \caption{ (Color online) Compressibility $\sqrt{t}\kappa $ (dimensionless) vs $\mu$  for $h=1.1$ at different values of $t$.   At the critical point $\mu_{c3}$, there  is a background compressibility. The curves truly intersect at a single point after the background is removed.
} \label{fig:5} \end{figure}

Finally, we would like to mention that similar plots can be constructed  for compressibility across all phase boundaries,   see Fig.~\ref{fig:5}.  For compressibility, it is important (as stressed before) to include all higher order terms in Eq.(\ref{SOE}), as they are all singular at criticality; and such inclusion can be achieved efficiently by taking derivatives of Eq (\ref{density-EOS}) with respect to $\tilde{\mu}$ together with performing proper iteration via Eq.(\ref{Pres-p})-(\ref{dress-X}) .  The expression of the critical behavior near various phase boundaries are given by
 \begin{eqnarray}
V{\rm -}F: \,\, \tilde{\kappa} &\approx&- \frac{1}{2\sqrt{2\pi t}} {\rm Li}_{-\frac{1}{2}} \left(-e^{\frac{(\tilde{\mu}-\mu_{c1})}{t}}\right) \\
 F{\rm -}PP : \,\,  \tilde{\kappa} &\approx &\kappa_{o3} -\frac{\lambda_{4}}{\sqrt{t}} {\rm Li}_{-\frac{1}{2}} \left(-e^{\frac{2(\tilde{\mu}-\mu_{c3})}{t}}\right)\\
V{\rm -}P: \,\,  \tilde{\kappa}&=& -\frac{2}{\sqrt{\pi t}}
{\rm  Li}_{-\frac{1}{2} }\left(-e^{\frac{2(\tilde{\mu}-\mu_{c2})}{t}}\right)\label{comp:F-PP}\\
P{\rm -}PP:  \,\,\tilde{\kappa} &= & \kappa_{o4} -\frac{\lambda_{5}}{\sqrt{t}} {\rm Li}_{-\frac{1}{2}} \left(-e^{\frac{(\tilde{\mu}-\mu_{c4})}{t}}\right)
\end{eqnarray}
where $\kappa_{o3}, \kappa_{o4}, \lambda_{4}, \lambda_{5}$ are given
 \begin{eqnarray}
 \kappa_{o3} &=&\frac{1}{2\pi\sqrt{a}}, \,\,\lambda_{4}= \frac{2}{\sqrt{\pi}} \left( 1+ \frac{ \sqrt{a}}{2\pi} -\frac{a}{2\pi^2}\right)\nonumber\\
 \kappa_{o4}&=&  \frac{ 2}{\pi\sqrt{b}}\left( 1- \frac{3\sqrt{b}}{\pi} +\frac{6b}{\pi^2}\right),\nonumber \\
  \lambda_5 &=& \frac{1}{2\sqrt{2\pi}} \left( 1 +\frac{2\sqrt{b}}{\pi} -\frac{10 b}{\pi^2}\right).
\end{eqnarray}
The critical exponents $z=2$ and $\nu=1/2$ can be read off the universal scaling function 
\begin{equation}
 \kappa(\mu,T)=\kappa_0+ T^{\frac{d}{z}+1 -\frac{2}{\nu z}} {\cal F} \left(\frac{\mu-\mu_{c}}{ T^{\frac{1}{\nu z}}}\right)
\end{equation} 
with an universal scaling function ${\cal F}(x)={\rm Li}_{-\frac{1}{2}}(x)$.

\section{VI. Conclusion}

We have studied the quantum critical phenomena of strongly attractive 1D
Fermi gas via exact BA solution. We have obtained the equation of state with
high precision from zero temperature up to the temperature scale of binding
energy.  From the equation of state, we have obtain the exact scaling form
for density, compressibility, and spin susceptibility in the vicinity of the
$T=0$ phase boundaries between different phases. These scaling forms
illustrate the universal TLL signature and physical origin of quantum
criticality. The excitations near various phase boundary is such that their
critical behaviors are either described by that of free fermions and that
mixtures of fermions with mass m and 2m.

Our exact results can help analyze the recent experiments on attractive
1D spin-imbalanced atomic Fermi gas \cite{Hulet}. 
It can also help to verify the working of an algorithm for determining the
$T=0$ phase boundary and quantum critical behavior based solely on measured
density profile \cite{ZhouHo}.  The BA method proves to be a powerful tool to
obtain exact results for quantum critical behavior. Further application of
this method to other integrable systems including spinor gases may reveal
new excitations and interesting physics.

{\bf Acknowledgment.} This work is supported by NSF Grant DMR-0907366 and by DARPA under the Army Research Office Grant Nos. W911NF-07-1-0464, W911NF0710576. XWG  has been supported by the Australian Research Council. He acknowledges the Ohio State University for their kind hospitality.

\vspace{0.8cm} 

\appendix{\bf Appendix: Coefficients}

The pressures
(\ref{Pres-p}) and  (\ref{Pres-u}) provide the precise equation of
states for studying  two-component Fermi gases with population
imbalance.  By iteration we found the analytical equation of state
(\ref{SOE}) which  provides high insights into many-body effect in the so
called FFLO phase.  The first two coefficients  are given in (\ref{Y_1_2}).
The higher order correction terms  are given by
\begin{eqnarray}
&&Y_{\frac{3}{2}}^b=\frac{1}{\pi^{\frac{3}{2}}}\left[f_{-\frac{3}{2}}^b\left(
\frac{\sqrt{2}}{8}\left(f_{\frac{3}{2}}^b\right)^2 f_{\frac{3}{2}}^u
+\frac{1}{2}\left(f_{\frac{3}{2}}^u\right)^2
f_{\frac{3}{2}}^b+\frac{1}{48}\left(f_{\frac{3}{2}}^b\right)^3\right.\right.\nonumber\\&& \left.\left.
 +\frac{\sqrt{2}}{3}\left(f_{\frac{3}{2}}^u\right)^3\right)+f_{-\frac{1}{2}}^b\left(
\frac{3}{16}\left(f_{\frac{3}{2}}^b\right)^2 f_{\frac{1}{2}}^b
+\frac{1}{\sqrt{2}}\left(f_{\frac{3}{2}}^b\right)^2 f_{\frac{1}{2}}^u
\right.\right.\nonumber\\
&&
\left.\left.
+\frac{3}{2}\left(f_{\frac{3}{2}}^u\right)^2 f_{\frac{1}{2}}^b
+2f_{\frac{3}{2}}^bf_{\frac{3}{2}}^uf_{\frac{1}{2}}^u+\frac{3\sqrt{2}}{4}f_{\frac{3}{2}}^bf_{\frac{3}{2}}^uf_{\frac{1}{2}}^b\right)\right.\nonumber\\
&&\left.+f_{\frac{1}{2}}^b\left(
\frac{1}{\sqrt{2}}\left(f_{\frac{1}{2}}^b\right)^2 f_{\frac{3}{2}}^u
+\frac{1}{8}\left(f_{\frac{1}{2}}^b\right)^2 f_{\frac{3}{2}}^b
+\frac{1}{\sqrt{2}}\left(f_{\frac{3}{2}}^b\right)^2
f_{-\frac{1}{2}}^u\right.\right.\nonumber\\
&&
\left.\left.
+2f_{\frac{1}{2}}^bf_{\frac{1}{2}}^uf_{\frac{3}{2}}^u+\sqrt{2}f_{\frac{1}{2}}^bf_{\frac{1}{2}}^uf_{\frac{3}{2}}^b\right)\right]\nonumber\\
&&
-\frac{1}{\sqrt{\pi}}\left(\sqrt{2}f_{\frac{1}{2}}^b
f_{\frac{5}{2}}^u +\frac{1}{16} \left(f_{\frac{3}{2}}^b\right)^2 +
\frac{1}{\sqrt{2}}f_{\frac{3}{2}}^b
f_{\frac{3}{2}}^u\right),\nonumber \\
&&Y_{\frac{3}{2}}^u=\frac{1}{\pi^{\frac{3}{2}}}\left[\frac{1}{6}\left(f_{\frac{3}{2}}^b\right)^3f_{-\frac{3}{2}}^u+f_{-\frac{1}{2}}^b\left(
\frac{1}{8}\left(f_{\frac{3}{2}}^b\right)^2
f_{-\frac{1}{2}}^u\right.\right.\nonumber\\
&&
\left.\left.
+\left(f_{\frac{3}{2}}^b\right)^2 f_{\frac{1}{2}}^u
+\frac{1}{\sqrt{2}}f_{\frac{3}{2}}^u f_{\frac{3}{2}}^b
f_{\frac{1}{2}}^u\right)\right.\nonumber\\
&&\left.+f_{\frac{1}{2}}^b\left(
\frac{1}{2}\left(f_{\frac{3}{2}}^b\right)^2 f_{-\frac{1}{2}}^u
+\sqrt{2}\left(f_{\frac{1}{2}}^u\right)^2 f_{\frac{3}{2}}^b
+\frac{1}{\sqrt{2}}f_{\frac{1}{2}}^bf_{\frac{1}{2}}^uf_{\frac{3}{2}}^u\right.\right.\nonumber\\
&&\left.\left.
+\frac{1}{4}f_{\frac{1}{2}}^bf_{\frac{1}{2}}^uf_{\frac{3}{2}}^b+\sqrt{2}f_{\frac{3}{2}}^bf_{\frac{3}{2}}^uf_{-\frac{1}{2}}^u\right)\right]\nonumber\\
&&-\frac{1}{\sqrt{\pi}}\left(\frac{1}{2}f_{\frac{1}{2}}^u
f_{\frac{5}{2}}^b +f_{\frac{3}{2}}^b
f_{\frac{3}{2}}^u\right), \nonumber
\end{eqnarray}
which reveal  cluster-cluster interacting   effect.


\begin{thebibliography}{99}
\bibitem{footnote} Our approach is different from that of Ref. \cite{Orso}.  The equation (8) in Ref. \cite{Orso}  can be expressed as $\mu=-\frac{1}{2}-\frac{1}{2\pi}\left[2Q-\left(2\mu+2h+1)\tan^{-1}(2Q)\right)\right]$ with $Q=\sqrt{(\mu+h)/2}$ in dimensionless unit, which is the same as our result (\ref{F-PP}) via a rescaling $h\to h/2$.
\bibitem{note}  The equations in Ref.\cite{Erhai} is obtained by (i) approximating the $[...]$ in Eq.(\ref{Pres-p}) and (\ref{Pres-u}) by 1, (ii) omitting the $t^{3/2}$ correction terms in Eq.(\ref{dress-X}),
  and (iii) by taking the $Tc\rightarrow \infty$  in string equations \cite{Takahashi}.
\end{thebibliography}
\end{document}